# Senior Project Management System: Requirements, Specification, and Design Issues

Ahmed Al-Dallal
Department of Computer Engineering
Kuwait University
P.O. Box 5969, Safat 13060, Kuwait

***Abstract*—Senior project is a typical essential course in computing educational programs. The course involves the selection of a project problem, the submission of various documents, and intensive communication among the project team members and between them and the course instructors. To facilitate all these tasks, we introduce the senior project management system (SPMS) that organizes and manages previous, current, and proposed senior projects in all of their stages along with proper ways of communication between the students and course instructors. This paper explains the system requirements and specifications and discusses related design issues. The paper shows the importance of well documenting the specifications and requirements of software systems and paying considerable attention to system design, which has a positive impact on implementing high quality systems.**

## I. INTRODUCTION

Computing-oriented degree programs typically include a senior project that involves developing a computer-based system [1, 2, 3, 4]. Senior project or capstone project courses traditionally span over many stages over many months, which is why developing a system to manage all the stages, will be very helpful for both students and instructors. Several systems have been developed for senior project management (e.g., [5, 6, 7, 8]). However, these systems do not assist students selecting their senior projects. In addition, the requirement and design issues related to the development of these systems are not discussed.

In this paper, we introduce a senior project management system called SPMS. The system manages previous, current, and proposed senior projects. The system is developed using a prototyped waterfall development model [9] and its specification and design are visualized using Unified Modelling Language (UML) diagrams [10]. Using this system, instructors place proposed projects in the system. Senior students are allowed to register in the system. Registered students can select their preferred projects from the proposed projects list or request to do their own project ideas, which become pending projects waiting for the instructor's approval. Once a project is selected, it becomes a current project. For a current project, the system allows the students to submit the intermediate and final reports. The instructor can use the system to view the submitted reports for a current project, write comments, and provide the grades. The system allows the students of a specific current project to view their submissions and view comments and grades. Once the project is completed, the instructor has to set the project as a previous project. This will help in building a database for the completed projects and will ensure that projects will not be repeated in the future. The system is open to the public such that any user can view the list and details of previous projects. However, only instructors can view the list and details of current projects. Both instructors and students can view proposed projects. The system is fairly simple and the dialogue boxes are self-explanatory. Users only need the basic knowledge needed to go through any other website they may use on a day-to-day basis, in order to be able to find their way around the system. The system offers students an external place, outside of university walls, to display their work in the hope of getting future opportunities. These opportunities will be in the form of funding from interested universities and businesses or ways of contact to help the scientific community further advance and manifest the ideas presented in the projects into services and products for the betterment of humanity.

The system has several advantages and benefits. By this system, all projects can be found in one place for anyone wishing to benefit and learn from previous projects while guaranteeing that no projects are repeated or stolen in the future. In addition, the system insures that university faculty members and students alike will have a simple format to follow and accomplish all stages of the senior project without any confusion that may arise. University faculty members and senior students are expected to have the necessary background to be able to operate and use the system with ease with the experience expected by their respective experiences.

The paper is organized as follows. Section 2 reviews related work. Section 3 describes the development of the user interface prototype. In Section 4, we discuss the requirement and specification issues. Design issues and models are discussed in Section 5. Possible future extensions to the system are discussed in section 6. Finally, Section 7 concludes the paper.

## II. Related Work

Several senior project management tools have been developed. Olarte et al. [5] proposed a tool that offers support for several features. The features include document management and a way of communication among students and between students and professors in the hope of building a learning community. Another feature is informing students of each other's progress to increase motivation and competitiveness, but what it does not offer is a database for previous projects, clear places for submission or a way for instructors to grade the project at the different stages. Olarte et al. did not discuss related requirements and design issues of the proposed system.

University of Pennsylvania [11] offers the option to upload capstone projects but only after they are done. This is helpful for having a reference for students in the future who wish to work on their senior projects. The students would avoid working on projects already implemented, but it allows them to learn from previous experiences. The system does not offer a way for students and instructors to communicate during the project or a place to submit the different stages of the projects. Students in ALHOSN University [6] developed a system that organizes the project stages and tasks while giving instructors the option to follow up on the progress of the different stages of the projects. Students can view project ideas, select the idea they prefer, and then they have to wait for their instructor's approval. Instructors set tasks and their deadlines where students can in return ask the instructor for further details in case something is not clear.

A web-based tool called Web Industrial Experience Resource [7] was introduced to 175 senior students in Monash University. The system provided many features such as templates and document samples so that students do not have to waste time and work on them from scratch. In addition, the features include an archive for accessing past projects to learn from past experiences, resources to help in researching recent technical subjects, a forum to discuss problems, and an option to store and backup files.

A similar system [8] was discussed that included features such as search and inquiry for students looking for previous capstone projects, monitoring and evaluation for instructors to decide deadlines and grades for each stage, and bulletin board and study note sections for communication purposes. There are other systems for general course management, but do not necessarily include capstone project courses (e.g., Blackboard Learn [12], online course system (OCS) [13], and Schoology [14]).

## III. User Interface Paper Prototype

The aim of the user interface paper prototype is to gather requirements using paper prototyping from various possible scenarios from the requirements specifications. A minimum of three scenarios can be conducted using this paper prototype. The paper prototype was tested by three users.

### A. *Scenarios*

The first scenario is viewing previous projects. Instructors, students, and public users open the SPMS main page and then select the "Previous Projects" button. Then the Previous Projects page opens with a list of the projects' names. When the user clicks on one of the projects, a new page opens with the group name, abstract, and description for the selected project. The second scenario is selecting a project. The student either selects the "Proposed Projects" button or the "New Idea" button. If the "Proposed Projects" button is selected, a new page opens with a list of project ideas suggested from the instructor. When the student selects a project idea, another page opens with the description. He then goes back and chooses the project idea he likes by clicking the empty box next to it. A check mark will appear and then to finalize the selection, he selects the "submit" button. The proposed project then becomes a current project. If the "New idea" button is selected, the student must write the description for the project and then select the "submit" button. The student waits for the instructor's approval so that the project can become a current project. The third scenario is submitting project stages. At each stage an upload option is there to upload the work/report for that stage with a "submit" button. The work will be graded by the instructor and he is also given the option to leave comments on the work done by the group for this specific stage of the project.

### B. *Interface Evaluation*

The interface evaluation activity was performed by the developer and three external evaluators. The objective of the user evaluation is to uncover any usability issues or any features lacking from the interface. In addition to the aforementioned steps, this activity could check phrasing and metaphors included in the interface.

The interface evaluation process includes preparation, execution, and observation. The preparation activities are run before the actual interface evaluation. These activities include building a prototype, writing the goals for each scenario on a separate index card, and giving roles to team members.

Four steps were followed to evaluate the prototype. The first step is briefing the user. In this step, the main objective and functions of the system were explained, and it was made sure that the user understood what was written in the summary. The second step is assigning the user one task. In this step, the prepared index cards were handed to several users. Each user got one card. We made sure that each user fully understood what was written in the card. The third step is observing the user do the task. The users were able to perform the given tasks with relative ease with two of them commenting that the system in some aspects resembles an existing system. The last step is repeating with the other tasks. More functions and tasks were given to the users while making sure that each user played the role of the computer in at least two of the tasks.

In several instances and on different pages there was not an option to go back to the previous page, such as going back to the "Instructor Home Page", or to cancel the action at hand, such as not going through with a proposed project in the "Editing Proposed Project Page". Therefore, we revised the prototype to ensure that the system pages are connected to each other properly. Many buttons were added such as, "back", "home", and "cancel" to address this issue.

## IV. SOFTWARE REQUIREMENT SPECIFICATION (SRS)

The SRS lists all the system requirements including functional, nonfunctional, and interface requirements along with system models. It also covers the system's main features. Typically, SRS is useful for the system developers, possible marketing staff, and testers. The SRS contains a general description of the system that includes product perspective, product features, user characteristics, constraints, and assumptions and dependencies. It also contains system requirements that include functional requirements, use case models, nonfunctional requirements, and interface requirements. It also contains system models.

### *A. General Description*

The system itself is independent and self-contained. However, some functionalities depend on the connection status to a web server responsible for database and synchronization services. This connection is required to allow users accessing the system wherever they are using an internet connection to read, upload, and download files. In terms of user/marketing requirements, there are three main product features. The first one is viewing previous projects. The second feature is selecting a project, where students may select one of the proposed projects from the instructor, or students may propose a new project idea and wait for the instructor's approval. Submitting project stages is the third feature, where students upload project stages to the system and instructors leave comments on each stage and write the grades.

Regarding user characteristics, the system is fairly simple and the dialogue boxes are self-explanatory. Users only need the basic knowledge needed to go through any other website they may use on a day to day basis in order to be able to find their way around the system. Users include course instructors of all ages and students in their senior year of college. As constraints, uploading and downloading large files may present a problem, especially with slow internet connections. There are several assumptions and dependencies. For example, the system should be operated on a dependable server that is web based with a large storage capacity of at least 10 TB dedicated for this particular system to be able to store large amounts of previous projects for public users to access and current project stages uploaded by students.

### *B. System Requirements*

The SPMS main functional requirements are provided in the use case diagrams given in Fig. 1. There are several functional requirements that are not captured by the use case diagram such as the options to return to previous pages using "back", "home", and "cancel" buttons. These are for instances such as going back to the "Instructor Home Page", or to cancel an action at hand, such as not going through with a proposed project in the "Editing Proposed Project Page".

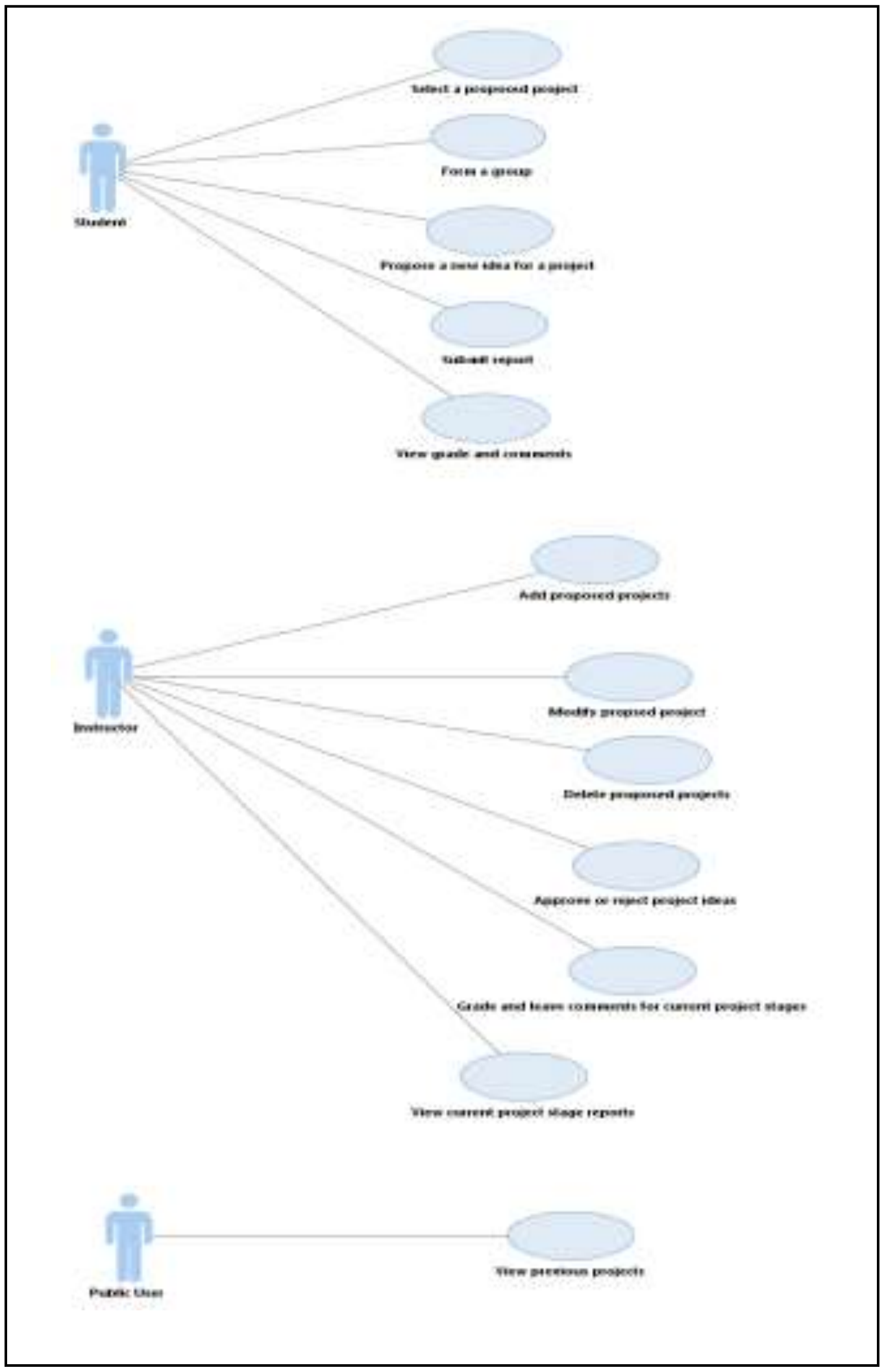

Figure 1. SPMS UML use case diagrams

We also considered several nonfunctional requirements for performance, availability, security, and portability purposes. For performance, the server should be able to load files at maximum time of 1 second per 5 MB. For availability, Functions like group formation, project proposing, and project selection should be available 99.9% of the time in the early stages of the senior project. All other functions of the system should also be available 99.9% in the remaining time of the project stages. In terms of security, only students and instructors with proper login information are permitted to access proposed and current projects. The system automatically logouts after 30 minutes of no activity as a security measure to prohibit unauthorized users to access an account in case an authorized user forgot to logout. Also, the system scans all files for viruses before being uploaded.

In terms of portability, the system components rely on a maximum of 10% of host-dependent code to simplify moving the system from one platform to another when needed. JavaScript is used as a proven portable language to implement the system's many functions.

Regarding interface requirements, the system is accessed online, which means no limitations or specific requirements are needed for certain hardware to run the system. Desktops, laptops, tablets, and smartphones will all work fine.

The system can interact with the software systems used in an educational institute's web servers. The communication with the system over the internet uses well known communication protocols such as HTTP. Grades and comments for stage reports can be typed using either keyboard or keypad. Stage files of any format can be uploaded. Project data such as instructor information, student information, and all related project information are stored in text files. The domain model of the system is provided in Fig. 2.

## V. SPMS High Level Design

The framework for the system is a department offering a computing program in an educational institute, instructors overseeing the senior projects, and senior students in the department. The development model used is the waterfall model that is plan-driven with separate and distinct phases of specification and development and all of the process activities were planned in advance. Uniform Modelling Language (UML) is the modeling language used for modeling the design components and their relationship. JavaScript is used to write the code for the system because the system is a web-based application which needs to be compatible with all browsers. Since the students need to upload stage files, the system must support all formats, such as PDF, Word Documents, JPEG, etc. All communications are done over the internet using TCP/IP and HTML protocols. The User Interface (UI) library and an existing JavaScript crypto API library for security are mainly used for programming the system.

The top level components of the system and their relationships are presented in this section. In addition, design alternatives and issues are discussed.

### A. *Component View*

The only hardware needed for the system is a dependable web server in an educational institute dedicated for this particular system to be able to store large amounts of previous projects for public users to access and students to upload and download large files. The server should be in the educational institute intending to use the system and not in any other establishment or country, in case any repairs or bug fixes are needed that require quick intervention. As software, the system uses Oracle database management system to manage the project, instructor, and student databases. We used Oracle because it is more suitable for servers than other database management systems such as Microsoft Access.

### B. *Structure, Interactions, and Dependencies*

The system functionalities depend on the status of the connection to the web server responsible for database and synchronization services, so that users could access it wherever they are using an internet connection to read, upload, and download files. Proposed projects and the list that includes all of them are connected to current projects and their list. The same applies to current projects and previous projects. A proposed project cannot become a previous project without passing through the current project stage.

### C. *Design Issues and Alternatives*

While designing the system, we considered several issues and alternatives as follows.

Issue #1: Is the system accessed using the internet or by a local area network?

Alternative A: The system is accessible using the internet.

Advantages: The system can be accessed wherever you are.

Disadvantages: Slow internet speeds may result in longer time to upload and download files.

Alternative B: The system is accessible using a local area network connecting the department's computers.

Advantages: It is faster to access the system's many functions and less time is needed to upload and download files.

Disadvantages: The system can only be accessed using the department's computers.

Issue #2: Is the code for the system written using JavaScript or C/C++?

Alternative A: The code for the system is written using JavaScript.
Advantages: Browsers support JavaScript.
Disadvantages: The code may have security gaps and therefore it requires strong security protocols and tools.
Alternative B: The code for the system is written using C/C++.
Advantages: Faster for standalone applications accessed via local area networks.
Disadvantages: C/C++ requires a compiler to operate and it is not suitable for web applications.

Issue #3: Projects run through different stages: proposed, current, and previous.

Alternative A: In the UML class diagram, only one class is used to represent the project, regardless of its state.

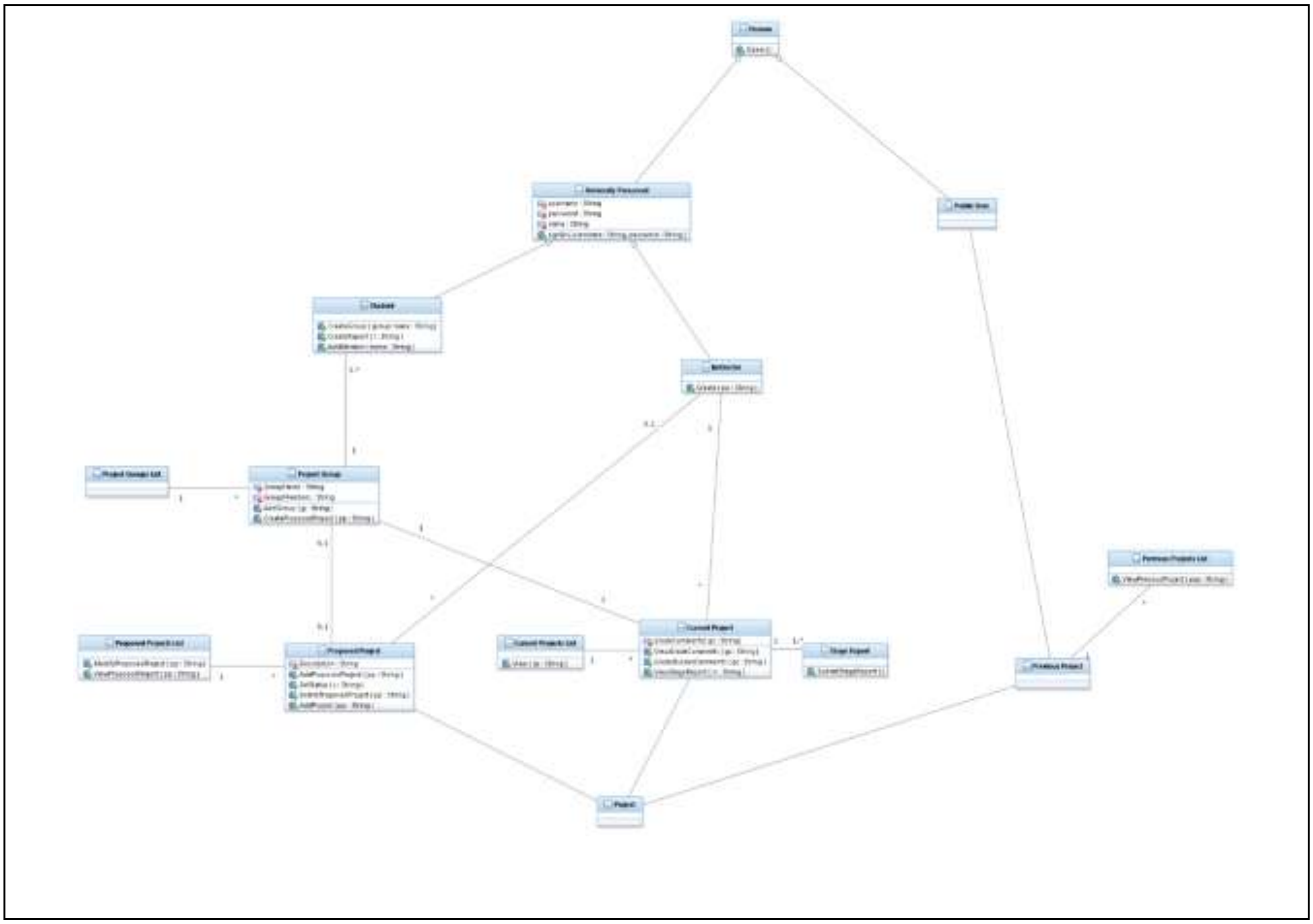

Figure 2. UML domain model

Advantages: Fewer requirements and fewer functions are needed.
Disadvantages: Attributes and functions are not distinct in some cases.
Alternative B: The project goes through different stages and, therefore, each stage has a distinct class in the UML class diagram. These classes are designed to be subclasses of a project superclass.
Advantages: Each step is clear, and unnecessary information is disposed once we move to the next stage.
Disadvantages: More work is done to satisfy the many requirements and functions for the system.

### D. *Selected Design and Justification*

We decided to have the system accessible using the internet. One of the most important requirements for the system is being able to access the system anywhere you go and not being limited to only one place for access, which ultimately renders the system useless. In terms of code, we decided to use JavaScript. The justification is that the system will run as a web-based application which needs to be compatible with all browsers. Regarding the states of a project, we decided to have the project go through different stages. The justification is that the web based server has a limited storage capacity and cannot store all information over many years for all students. After the project is completed, the project files submitted at different stages become useless and must be disposed.

## VI. Future Extensions

Many features can be added to the system in the future to make it more complete. The system could be used by instructors for other courses that have projects or for general course management. Instructors can make announcements, posts assignments, upload lecture notes, or anything needed to enrich and ease the learning process for students. Expanding the system into a mobile based application will also be a future addition since the market and usage for mobile applications is increasing. The idea is that the system can be used by universities worldwide to build a bigger database, instead of each university having to build a system from scratch and only be utilized for that certain university, resulting in student ideas not receiving the spotlight they deserve.

## VII. Conclusions

The system presented in this paper is essential to senior students in every university and is a convenient way to help them concentrate and focus on the project itself. Making sure everything related to the project is organized and secure is a big source of added stress when working on senior projects while having to worry about other courses. The hassle of having everything on papers or having to carry a laptop everywhere you go is unnecessary. The system can be accessed anywhere and at any time.